\documentclass[10pt]{article} %
\usepackage[preprint]{tmlr}

\usepackage{hyperref}
\usepackage{url}

\title{\sys: Bug-Property-Guided Program Mitigation}

\author{\name Han Zheng\textsuperscript{1}
  \email han.zheng@epfl.ch
  \AND
  \name Rafaila Galanopoulou\textsuperscript{1}
  \email rafaila.galanopoulou@epfl.ch
  \AND
  \name Ilia Shumailov\textsuperscript{2}
  \email ilia.shumailov@gmail.com
  \AND
  \name Tianqi Fan\textsuperscript{3}
  \email tqfan@google.com
  \AND
  \name Aiden Hall\textsuperscript{3}
  \email aidenhall@google.com
  \AND
  \name Dominik Maier\textsuperscript{3}
  \email dmnk@google.com
  \AND
  \name Mathias Payer\textsuperscript{1}
  \email mathias.payer@nebelwelt.net
  \\[0.8em]
  \addr{
    \textsuperscript{1}EPFL
    \hspace{3em}
    \textsuperscript{2}Meta
    \hspace{3em}
    \textsuperscript{3}Google
  }
}

\def\openreview{\url{https://openreview.net/forum?id=XXXX}} %

\usepackage{hyperref}
\usepackage{tikz}
\graphicspath{{../../}}
\usepackage{amsmath}

\usepackage{filecontents}

\usepackage{multirow}
\usepackage{xcolor}
\usepackage{booktabs}
\usepackage{listings}
\usepackage{xspace}

\usepackage{enumitem,kantlipsum}

\usepackage{pifont}
\newcommand{\xmark}{\ding{55}}%
\newcommand{\cmark}{\ding{51}}%
\def\halfcheckmark{\tikz\draw[scale=0.4,fill=black](0,.35) -- (.25,0) -- (1,.7) -- (.25,.15) -- cycle (0.75,0.2) -- (0.77,0.2)  -- (0.6,0.7) -- cycle;}

\newcommand{\ie}{\textit{i.e.,} }
\newcommand{\eg}{\textit{e.g.,} }

\newcommand{\sys}{\textsc{CodeMechanic}\xspace}
\newcommand{\sysfast}{\textsc{CodeMechanic$_{1}$}\xspace}
\newcommand{\sysfull}{\textsc{CodeMechanic}\xspace}

\newcommand{\patchagent}{\textsc{PatchAgent}\xspace}

\newcommand{\patchfast}{\textsc{PatchAgent$_{1}$}\xspace}
\newcommand{\patchfull}{\textsc{PatchAgent}\xspace}

\newcommand{\sanpatch}{\textsc{San2Patch}\xspace}
\newcommand{\sanfast}{\textsc{San2Patch$_{1}$}\xspace}
\newcommand{\sanfull}{\textsc{San2Patch}\xspace}

\newcommand{\keyvar}{\textit{key variables}\xspace}
\newcommand{\contribone}{\textit{Bug-Property-Guided Mitigation}\xspace}
\newcommand{\contribtwo}{\textit{Two-Dimensional Context Extractor}\xspace}
\newcommand{\contribthree}{\textit{In-Prompt Human Knowledge}\xspace}

\definecolor{pinkshocking}{rgb}{0.99,0.01,0.84}

\usepackage{listings}
\usepackage{subcaption}

\definecolor{codegreen}{rgb}{0,0.6,0}
\definecolor{codegray}{rgb}{0.5,0.5,0.5}
\definecolor{codepurple}{rgb}{0.58,0,0.82}
\definecolor{backcolour}{rgb}{0.95,0.95,0.92}

\lstdefinestyle{mystyle}{
    backgroundcolor=\color{backcolour},
    commentstyle=\color{codegreen},
    keywordstyle=\color{magenta},
    numberstyle=\tiny\color{codegray},
    stringstyle=\color{codepurple},
    basicstyle=\ttfamily\footnotesize,
    breakatwhitespace=false,
    breaklines=true,
    captionpos=b,
    keepspaces=true,
    numbers=left,
    numbersep=5pt,
    showspaces=false,
    showstringspaces=false,
    showtabs=false,
    tabsize=2
}
\lstdefinestyle{cpp}{
    language=C++,
    basicstyle=\ttfamily\footnotesize,
    keywordstyle=\color{blue}\ttfamily\footnotesize,
    stringstyle=\color{red}\ttfamily\footnotesize,
    commentstyle=\color{codegreen}\ttfamily\footnotesize,
    morecomment=[l][\color{magenta}\footnotesize]{\#}
}

\usepackage{framed}

\definecolor{lightgray}{gray}{0.95}
\definecolor{darkgray}{rgb}{0.6, 0.6, 0.6}

\definecolor{myyellow}{HTML}{fff2cc}
\definecolor{mygreen}{HTML}{d9ead3}
\definecolor{myblue}{HTML}{c9daf8}
\definecolor{myred}{HTML}{f4cccc}

\newcommand{\greensquare}{\textcolor{mygreen}{\rule{1.5ex}{1.5ex}\xspace}}
\newcommand{\bluesquare}{\textcolor{myblue}{\rule{1.5ex}{1.5ex}\xspace}}

\lstdefinestyle{stacktrace}{
  backgroundcolor=\color{lightgray},
  basicstyle=\ttfamily\small,
  breaklines=true,
  frame=single,
  columns=fullflexible,
  showstringspaces=false
}

\newenvironment{formal}{%
  \MakeFramed{\advance\hsize-\width\FrameRestore}%
  \noindent\hspace{-4.55pt}%
  \begin{list}{}%
    {\setlength\leftmargin{0pt}%
     \setlength\rightmargin{7pt}}%
  \item\relax
  \vspace{2pt}%
}
{%
  \vspace{2pt}%
  \end{list}%
  \endMakeFramed%
}

\newcommand{\numarvocompare}{$101$\xspace}

\begin{document}

\maketitle

\begin{abstract}
Automated testing discovers vulnerabilities faster than developers can
investigate and repair them, leaving an interval in which known memory
corruptions remain exploitable. End-to-end LLM repair agents can shorten this
interval, but they synthesize open-ended code changes and commonly validate them
only by replaying a proof of concept (PoC). This weak oracle accepts patches that
silence the observed crash by changing unrelated behavior, making unintended
deployment risky.

We present \sys, a bug-property-guided system for generating constrained
mitigations for spatial memory corruption. Instead of asking an LLM to generate
a permanent repair, \sys reconstructs the
violated memory-safety property from the crash, validates the dereferenced
pointer and its buffer range, and inserts a local fail-stop guard before the
dangerous access. The guard terminates execution when the boundary check fails.
The resulting mitigation deliberately trades availability for
security: it can convert potential remote code execution into controlled
termination while developers investigate the root cause and prepare a permanent
repair. \sys combines a two-dimensional static and dynamic context extractor with
in-prompt debugging knowledge and stepwise validation to limit the effect of LLM
errors. On 101 real-world ARVO bugs, the first attempt of \sys produces 47.6\%
more plausible patches (i.e., patches that pass PoC-replay validation) than the best baseline while using 91\% fewer
tokens. Manual audit further shows that \sys produces 3.4$\times$--4.3$\times$
more patches semantically equivalent to developer-written repairs.
\end{abstract}

\section{Introduction}
\label{sec:intro}

Modern software systems add functionalities and thereby increase 
complexity~\citep{dullien2018security}. 
However, this growing complexity exceeds human comprehension limits, 
developers often misinterpret abstractions and struggle to reason 
across diverse interfaces, 
ultimately leading to the emergence of software vulnerabilities~\citep{chrome_checklist}.
On Bugcrowd, a third-party bug bounty platform, hundreds of thousands of vulnerabilities 
were reported and acknowledged by vendors~\citep{bugcrowd_report}, 
in addition to those identified and addressed internally. 
These vulnerabilities are not theoretical~\citep{chrome_checklist}. 
In contrast, 
they can be, and have been, exploited by skilled 
attackers~\citep{zeroclick_pixel,zeroclick_imessage,oneclick_android}. 
Proactively discovering and remediating vulnerabilities before they are 
weaponized is therefore essential to ensuring user safety.

Over the past decade, fuzzers and static analyzers have 
discovered an astonishing amount of
vulnerabilities~\citep{afl,fioraldi2020afl++,libfuzzer,codeql}. 
However, this rapid growth introduced a new challenge: 
the volume of bugs now exceeds the capacity 
of developers to triage and 
resolve them~\citep{too_many_bugs_fix}. For example, the Linux kernel currently 
has around 1,500 unresolved bugs reported by syzbot~\citep{syzbot}. 
Among these, memory corruption issues are the most exploited by 
attackers in real-world scenarios~\citep{memory_corruption_exploitable}.
This growing backlog highlights the urgent need for Automated Vulnerability 
Repair (AVR) techniques to reduce manual effort and help developers 
keep pace, especially with memory corruption vulnerabilities. It also motivates
rapid mitigations that reduce exposure while a permanent repair is developed.

The rapid advancement of Large Language Models (LLMs)~\citep{comanici2025gemini,chatgpt,liu2024deepseek}
has created new opportunities for automated program repair. A growing number of
AVR agents ask an LLM to localize a defect and generate a permanent
repair~\citep{tufano2019empirical,chen2019sequencer,feng2020codebert,huang2025template,xia2022less,xia2024automated,zhang2024fixing,huang2023empirical,PatchAgent}.
For real-world vulnerabilities, however, an unrestricted LLM-generated repair is risky: the LLM can modify arbitrary code, while the available oracle
often checks only that one PoC no longer crashes. An incorrect patch may
therefore pass validation while silently corrupting functionality or introducing
a new vulnerability. This mismatch makes open-ended patches difficult to deploy
without a human in the loop.
Despite its improvement, several key challenges limit  
their broader applicability:

\textbf{(C1) Unbounded effects of LLM-generated patches.}
LLMs are probabilistic and can generate arbitrary ``plausible'' patches.
With only a PoC-based oracle, deleting a code path, corrupting a fuzz driver, and
repairing the underlying bug may all appear successful. For unattended
mitigation, the patch space must instead bound the possible side effects and
make failure behavior predictable.

\textbf{(C2) Insufficient context for reconstructing a memory-safety property.}
The crash site alone rarely contains the pointer, object base, and object end
needed to state the violated memory-safety condition. Without both code and
runtime context, the LLM may hallucinate a boundary or overfit to the observed
execution.

\textbf{(C3) Dependence on expert debugging knowledge.}
Recovering the relevant pointer and object bounds requires deliberate use of
code search, macro resolution, GDB, and sanitizer metadata. Without guidance,
an LLM may invoke these tools repeatedly or misinterpret their errors.

To address these challenges, we propose the 
following solutions:

\textbf{(S1) Bug-Property-Guided Mitigation}.
We constrain the output to a local guard derived from the spatial memory-safety
property. The LLM identifies the dereferenced pointer and its valid buffer range,
and \sys inserts a check such as
\lstinline{if (!valid_pointer(p)) exit(0);} immediately before the dangerous
access. This is intentionally a temporary mitigation rather than a permanent repair. If the inferred condition is
violated, the guard terminates execution before the memory corruption can be
exploited. Restricting the patch format and location also prevents the
LLM from freely modifying unrelated program logic.
We call this constrained, temporary code change a \emph{mitigation patch}.
When the distinction from a permanent repair is clear, we use \emph{patch}
as shorthand.

\textbf{(S2) Two-Dimensional Context Provision}.
To overcome the challenge of insufficient bug context, 
we provide the LLM with a dual perspective: a code view and a 
data view.
In the code view, we supply macro definitions and representative 
examples of variable usage within the program, helping the LLM 
situate the bug in its broader contexts.
In the data view, we expose the LLM to a live debugger session 
paused at the crash site. This includes collected variable values 
and types, along with Sanitizer metadata~\citep{asan_debugger}, which 
enable the LLM to better comprehend the memory layout and underlying 
defect condition.

\textbf{(S3) Embedding Human Debugging Knowledge}.
Finally, to address the dependence on expert tool usage, we incorporate 
human debugging strategies directly into the prompts. These embedded 
guidelines offer a structured workflow for error handling. 
For example, prompts regarding global-buffer-overflow bugs require 
the LLM to validate the buffer name provided in the ASan log. 
Similarly, when encountering a symbol undefined error, the prompt 
instructs the LLM to call \lstinline{find_macro_definition} in order 
to resolve the missing symbol.

We implemented our prototype, \sys, and evaluated it against the 
state-of-the-art AVR agents, \patchagent and \sanpatch, 
on real-world vulnerabilities. 
The results show that \sys generates 47.6\% more plausible 
patches than 
the best competitor, while spending only 9\% of the tokens. 
With the default five-attempt setup, \sys achieves 13.3\% more plausible patches 
compared to the best competitor, with only 10\% of the cost. 
Through manual auditing of the plausible patches, 
we found that \sys produces 240\% and 325\% more 
semantically equivalent patches (functionally identical to the 
developer-written repairs) compared to \patchagent and \sanfull respectively.
This result indicates that constraining the patch can also be a 
strategy when developers repair the bugs.
Our contributions are:
\begin{itemize}
  \item We formulate bug-property-guided mitigation as a constrained
  alternative to open-ended LLM repair for spatial memory corruption.
  \item We introduce \contribone, supported by \contribtwo and
  \contribthree, to recover and validate a violated boundary property before
  generating a local guard.
  \item We implement and evaluate \sys against two state-of-the-art AVR agents,
  measuring plausible-patch count, cost, semantic equivalence, and failure modes.
  \item We will release \sys to support open science upon acceptance. %

\end{itemize}

\section{Background}
\label{sec:background}

\subsection{Exploitable Memory Corruption Bugs}
\label{ssec:background-memory}

Despite over three decades of academic and industry efforts~\citep{hack_stack1996}, 
memory corruption remains one of the most critical classes of software vulnerabilities. 
For instance, in Chromium, over 70\% of high-severity security issues are 
linked to memory safety bugs~\citep{memory_safety_chrome}. 
These vulnerabilities are widespread and are actively exploited by skilled attackers,
posing significant risks to end users.

Broad mitigations such as control-flow integrity, hardened memory allocators,
and sandboxing~\citep{harden_allocator,wahbe1993efficient,carlini2015control}
increase the cost of exploitation, but skilled attackers continue to bypass
them~\citep{chrome_full_rce,miracle_ptr_bypass,pwn2own25berling}. A permanent
repair remains the desired endpoint after a vulnerability is
reported. However, producing and validating that repair takes time. A targeted
local guard derived from the crash can block the known dangerous access while
developers prepare a permanent repair.

Over the past decade, the success of automated bug-finding techniques has led to the 
discovery of a massive number of security issues~\citep{afl,fioraldi2020afl++,libfuzzer,ossfuzz}. 
Unfortunately, the number of discovered bugs with potential security impact 
is growing faster than developers can address them~\citep{too_many_bugs_fix}. 
For example, in the Linux kernel alone, over 1,500 discovered bugs remain 
unfixed~\citep{syzbot}, 
and more than 20\% of reported bugs remain unresolved even one year after their initial 
report~\citep{syzbot_num_vm}.
This situation calls for both AVR techniques that assist developers and
constrained mitigations that provide immediate protection.
\begin{table}[t!]
\resizebox{\textwidth}{!} {
\begin{tabular}{l|cc|cc|c}
\toprule
                           & \multicolumn{2}{c|}{Patch Control}                          & \multicolumn{2}{c|}{Repairing Context} & \multicolumn{1}{c}{\multirow{2}{*}{\begin{tabular}[c]{@{}c@{}}Debugging \\ Knowledge\end{tabular}}} \\
                           & Localization      & \multicolumn{1}{c|}{Format} & Code View          & Data View         & \multicolumn{1}{c}{}                                                                                \\ \midrule
Constraint Repairing~\citep{zhang2022program,gao2021beyond} & Manual & Constraint$^{*}$                         & \xmark                 & \cmark               & -                                                                                                 \\
Sanitizer-based agents~\citep{PatchAgent,kim2025logs}  & Any            & Any                         & \cmark                & \xmark                & \xmark                                                                  \\ \midrule                                
\sys              & CrashSite              & Constraint                        & \cmark                & \cmark               & \cmark                                                                                                 \\ \midrule
\end{tabular}
}
\caption{Comparision between \sys and AVR tools. *: based on patch template and may not compile.}
\label{tab:bg-compare}
\end{table}

\subsection{Automated Vulnerability Repair and Mitigation}
\label{ssec:background-repair}

LLMs have significantly improved automated repair
techniques~\citep{tufano2019empirical,chen2019sequencer,feng2020codebert,huang2025template,xia2022less,xia2024automated},
leading to several LLM-based AVR agents~\citep{zhang2024fixing,kim2025logs,PatchAgent}.
These agents typically begin with a PoC and a reproducible environment, use
execution traces or sanitizer reports to localize the bug, and ask the LLM to
generate a permanent repair. This end-to-end objective leaves the output
space open. Limited context or reasoning~\citep{qi2015analysis} can therefore
produce a patch that passes PoC replay by modifying unrelated functionality.
Without a comprehensive functional oracle, such a patch still requires human
review before deployment.

Constraint-based repair techniques reduce this freedom by enforcing structured
constraints and collecting variables through fuzzing or concolic
execution~\citep{zhang2022program,shariffdeen2021concolic,gao2021beyond}.
However, these approaches assume an explicit bug location, which is often
unavailable in practice, and rely on dynamic traces without source-level
comprehension. Their templates may also fail to compile when the selected
expressions do not fit the local program context.

\sys occupies the space between open-ended LLM repair and template-only
constraint repair, as summarized in~\autoref{tab:bg-compare}. It uses an LLM to
recover source-level expressions, but restricts the output to a boundary guard
at the crash site. \contribtwo combines code and data context, while
\contribthree guides tool use and error handling. \contribone then validates the
intermediate boundary property and emits a temporary mitigation. The objective
is rapid mitigation with bounded side effects, not automatic generation of a
permanent repair.

While this pattern can in principle extend to a bug class with a reliable
runtime oracle, this work focuses on spatial memory corruptions, for which a
pointer and object range expose a concrete violation condition. We exclude
temporal memory bugs because comparable object-liveness oracles are rarely
available in production environments.

\section{Motivation and Problem Definition}
\label{sec:motivation}

We now define the gap addressed by \sys. The goal is not to generate a permanent
repair. Given a reproducible crash, \sys instead
constructs a narrow guard that prevents the observed unsafe memory access
with predictable termination.

\subsection{Motivation}
\label{sub:motiv}

Current agents often optimize for passing validation rather than preserving the
program's intended logic~\citep{nong2025appatch}. When validation consists of
replaying one PoC, an arbitrary functionality change can be indistinguishable
from a permanent repair.

\begin{figure}[t!]
\begin{lstlisting}[style=cpp]
--- a/libavc/fuzzer/svc_dec_fuzzer.cpp
+++ b/libavc/fuzzer/svc_dec_fuzzer.cpp
@@ -304,7 +304,7 @@ void Codec::allocFrame()
 for(UWORD32 i = 0; i < num_bufs; i++)
 {
-   mOutBufHandle.u4_min_out_buf_size[i] = sizes[i];
+   sizes[i] = (mWidth / 2) * (mHeight / 2);
    mOutBufHandle.pu1_bufs[i] = (UWORD8 *) iv_aligned_malloc(NULL, 16, sizes[i]);
 }
\end{lstlisting}
\caption{A ``plausible'' patch generated by an AVR agent~\citep{PatchAgent}. It prevents the observed failure by changing the fuzz driver rather than the vulnerable library.}
\label{list:bad-fix-example}
\end{figure}

\textbf{Challenge 1: Unbounded patch effects (C1).}
Current AVR agents do not constrain the format or location of generated patches.
Existing benchmarks also lack comprehensive functionality tests, allowing
irrelevant or incorrect changes to pass validation as ``plausible.''
For example, \autoref{list:bad-fix-example} illustrates vulnerability 
OSS-VUL-65057, where a heap-buffer-overflow occurs in 
\lstinline{decoder/ih264d_format_conv.c}. State-of-the-art AVR agent~\citep{PatchAgent} 
incorrectly identifies the root cause in the fuzz driver and changes its buffer
allocation behavior. The PoC no longer exposes the original failure, but the
vulnerable library remains unchanged. Such ``plausible but incorrect'' outputs
show why an automatic response should be constrained by a security property
rather than entrusted with arbitrary program semantics.

\begin{figure}[t!]
\begin{lstlisting}[style=cpp]
4975: int htmlParseDocument(htmlParserCtxtPtr ctxt) {
// ...
5068:     if (CUR == 0)       // <= Crash Site
5069:   htmlAutoCloseOnEnd(ctxt);
// ...
5088: } // To this line, CUR do not show up
\end{lstlisting}
\caption{The crash-site snippet does not expose the pointer bounds needed to construct a mitigation.}
\label{list:bad-fix-example-2}
\end{figure}

\textbf{Challenge 2: Insufficient context for property reconstruction (C2).}
LLMs require sufficient code and runtime context to reconstruct the violated
boundary.
For memory corruption vulnerabilities, 
the model should be provided with either static code snippets and  
runtime data that include expressions \lstinline{buffer}, \lstinline{buffer_end},
and \lstinline{pointer}. 
Without this context, the LLM may hallucinate a boundary or overfit the
observed crash.
\autoref{list:bad-fix-example-2} illustrates this issue. 
Here, a global buffer overflow occurs because \lstinline{CUR} is a macro 
that expands to the dereferenced pointer \lstinline{ctxt->input->cur}. 
And this pointer should fall within the bounds of 
\lstinline{[ctxt->input->begin, ctxt->input->end)}. 
However, if an AVR agent simplify dumps the code snippet around the crash site,
even if it includes 100 lines of code, the necessary contextual information 
(\ie \lstinline{ctxt->input->begin} and \lstinline{ctxt->input->end} expressions)
remains absent. Consequently, the LLM lacks enough context to construct a guard
from the local snippet alone.

\textbf{Challenge 3: Missing human debugging knowledge (C3).}
Even when AVR agents provide both dynamic and static analysis tools, LLMs do
not inherently possess the knowledge of how to invoke these tools effectively
to reason about the root cause of a bug.  Instead, they often resort to
repetitive tool usage without making meaningful progress.  For example, in
\autoref{list:bad-fix-example-2}, the agent is equipped with tools such as
\lstinline{print_local_vars()}, \lstinline{print_var_value()}, and
\lstinline{find_macro_def()}. However, the LLM repeatedly invokes
\lstinline{print_local_vars()} despite receiving identical results each time.
Furthermore, when attempting to inspect the value of \lstinline{CUR} using
\lstinline{print_var_value()}, the LLM encounters a \textit{symbol not found}
error but fails to infer that \lstinline{CUR} is a macro. Consequently, it does
not proceed to call \lstinline{find_macro_def()} to retrieve the actual
definition. This illustrates why the mitigation pipeline must encode debugging
expertise: the LLM needs guidance to select the appropriate tools, interpret
errors, and recover a boundary that can be checked independently.

To address these challenges, \sys uses an explicit spatial memory-safety
property to restrict the patch space through \contribone
(\autoref{ssec:design-repair}). \contribtwo
(\autoref{ssec:design-context}) recovers the static and dynamic expressions
needed to instantiate that property, while \contribthree
(\autoref{ssec:design-knowledge}) guides tool use and error recovery. Together,
these components analyze a concrete crash and produce a local mitigation
whose behavior is easier to validate than an unrestricted LLM-generated repair.

\subsection{Bug Property Abstraction}
\label{sec:bug-abstract}

To formalize the mitigation contract of \sys, we first recap the relevant
property of spatial memory corruption and then define the condition enforced by
the generated guard.

\noindent \textbf{Bug Properties.} Memory safety requires that every pointer dereference remains within the bounds of a valid, live object~\citep{Payer18SS3P}. A violation of this property constitutes memory corruption: exceeding object bounds causes \emph{spatial} memory corruption, while referencing a deallocated object causes \emph{temporal} memory corruption. \sys focuses exclusively on spatial memory corruption, as temporal errors lack stable detection oracles in practice. Verifying object liveness typically requires smart pointer APIs~\citep{weakptr} or sanitizer debugging interfaces~\citep{asan_debugger}, neither of which are commonly available in open-source projects.

\noindent \textbf{Bug Model.} We model spatial memory bugs with the following boundary violation condition:
\begin{equation}
   (ptr < buf\_begin \; || \; ptr + sizeof(*ptr) > buf\_end) \label{eq:bound-check}
\end{equation}
where \lstinline{buf_begin} and \lstinline{buf_end} denote the start and end addresses (exclusive) of a valid memory buffer. A bug occurs when a pointer \lstinline{ptr} violates this condition and is subsequently dereferenced.

\noindent \textbf{Mitigation Contract.} The generated guard is inserted before
the pointer dereference and enforces \autoref{eq:bound-check}. If the condition
is violated, it terminates execution before memory is corrupted. This
response deliberately converts the dangerous execution into a denial
of service. It does not claim to provide a complete permanent repair.
Synthesizing the guard requires a minimal set of
\textbf{key variables}: the dereferenced \textit{pointer} and the
\textit{buffer range} defined by its base and end addresses. \sys retrieves
these expressions together with their type, size, and aliasing information and
uses them to construct a targeted mitigation.

\section{Design of \sys}
\label{sec:sys-design}

\begin{figure*}[t]
	\centering
	\includegraphics[width=\linewidth]{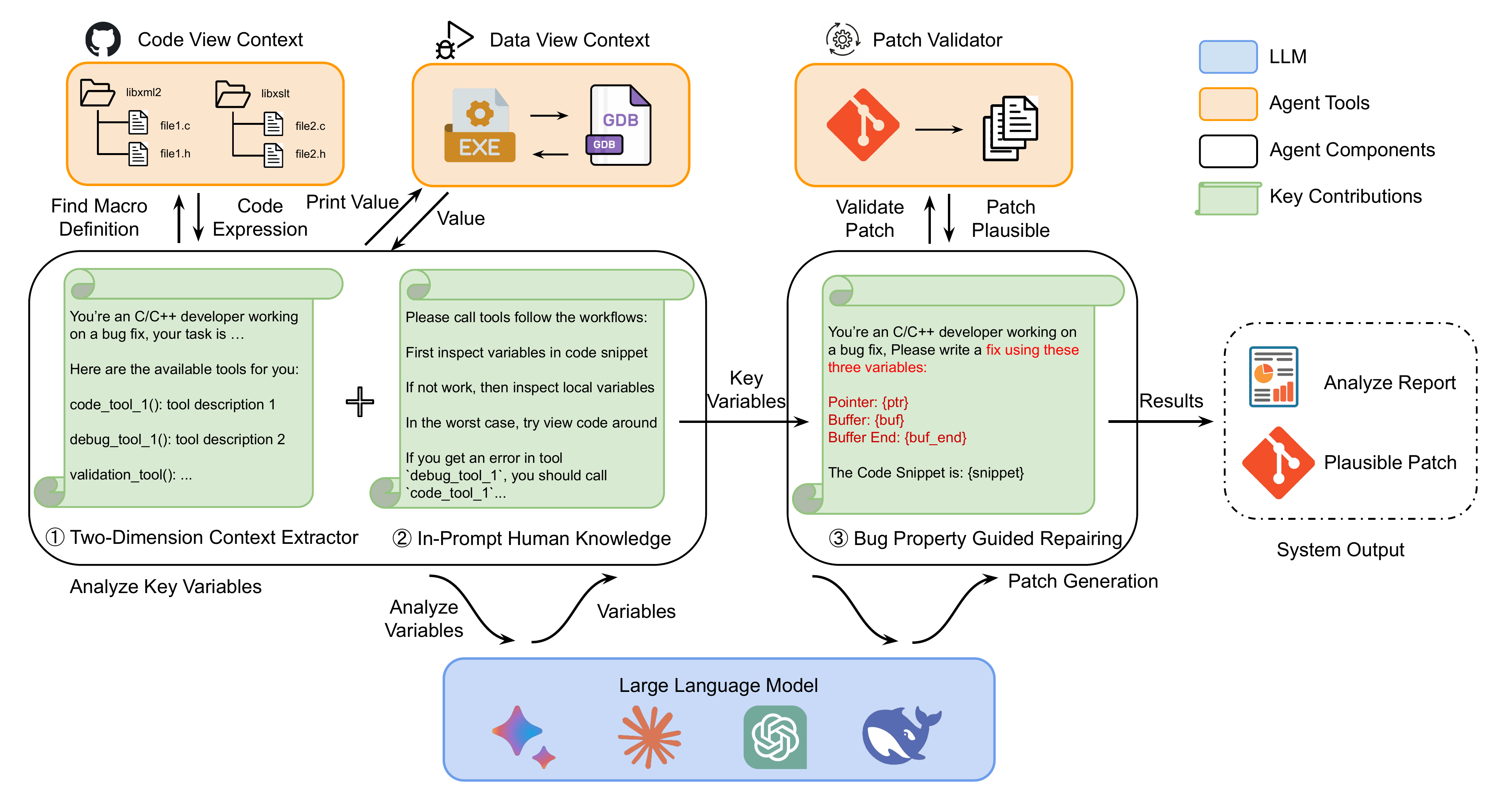}
	\caption{\sys's design. \contribtwo leverages both 
  code view contexts and data view contexts.}
	\label{fig:sys-design}
\end{figure*}

We present the design of \sys in~\autoref{fig:sys-design}. \contribtwo
provides the LLM with code and data views, and \contribthree guides tool calls
and error handling. These components recover candidate expressions for the
pointer and object range. \contribone validates these intermediate results and
uses them to instantiate the restricted guard. This decomposition
keeps the LLM responsible for source-level inference while using explicit
oracles to check the bug property on which the mitigation depends.

\subsection{Two-Dimensional Context Extractor}
\label{ssec:design-context}

To recover the expressions required by the mitigation contract, \sys introduces
a two-dimensional context extractor that integrates code-view and data-view
information.

\noindent \textbf{Code View Context.} In the code view, we provide three tools to help the LLM extract relevant diagnostic context from the codebase.

The \textit{code exploration} tool allows the LLM to inspect code beyond the immediate
crash site, improving its understanding of the broader context.

The \textit{macro resolution} clarifies symbols that may obscure the root cause
of a crash.
It resolves macro symbols to their underlying expansions using a simple language server,
enabling the LLM to reason about actual variables and operations. 
For example, in \autoref{list:bad-fix-example-2}, 
\textit{macro resolution} locates the definition of \lstinline{CUR} and 
expands it to \lstinline{*ctxt->input->cur}, 
revealing that the dereference involves a pointer.

The \textit{buffer inference} tool addresses the challenge that buffer boundaries 
relevant to a crash may be defined far from the crash site or even in separate components. 
It applies a heuristic based on the pointer expression's operands: 
specifically, it strips the trailing member access or index, 
and uses the remaining prefix as a pattern to search for related buffer 
references across the codebase, 
allowing the LLM to infer key variables needed for the boundary guard.
For example, in \autoref{list:bad-fix-example-2}, \textit{buffer inference} 
strips the trailing field access from \lstinline{ctxt->input->cur} 
and searches for \lstinline{ctxt->input}, 
surfacing references such as \lstinline{ctxt->input->base} 
and \lstinline{ctxt->input->end} as potential buffer boundaries.

\noindent \textbf{Data View Context.} In the data view, 
we provide three tools to collect and validate runtime context.

The \textit{pointer and buffer validation (with ASan metadata)} tool leverages 
ASan metadata to deterministically verify pointer values and buffer ranges, 
rather than relying solely on the LLM’s reasoning.
It cross-checks actual pointer values against sanitizer-reported error addresses,
and validates buffer ranges against memory layout metadata, ensuring that 
reported buffers lie within valid bounds. 
For example, an ASan report of \textit{global-buffer-overflow on address 
\lstinline{0x0000071da0a8}} provides both the fault address for pointer validation
and a region descriptor such as \textit{8-byte region 
[\lstinline{0x6020000bc410}, \lstinline{0x6020000bc418})]} for range validation.

The \textit{pointer and buffer validation (without ASan metadata)} tool 
handles stack-based buffer overflows, where ASan does not expose precise bounds
metadata.
It records the faulting address, resumes the crashing frame in GDB to measure 
the pointer's declared span, and scans local variables and arguments 
to identify which buffer range contains the faulting address. 
If no exact buffer-range match is found, it falls back to the candidate 
whose end lies closest to the fault address, 
ensuring reliable handling even without precise ASan metadata.

The \textit{pointer type analysis} tool addresses the subtlety that a pointer may be within bounds yet its dereference still causes an overflow, depending on the pointed-to type. It supplies type information alongside raw pointer expressions, enabling the LLM to correctly reason about the number of bytes accessed during a dereference. For example, a pointer \lstinline{ptr} of type \lstinline{size_t *} positioned at the end of a buffer will access 8 bytes upon dereference, silently overflowing the buffer even though \lstinline{ptr} itself appears in-bounds.

\subsection{In-Prompt Human Knowledge}
\label{ssec:design-knowledge}

Even with rich context, LLMs may struggle to use them effectively 
without domain knowledge, 
\eg calling tools in an ineffective order or misinterpreting error messages. 
To address this, \sys embeds human debugging knowledge 
into the prompt via few-shot prompting~\citep{touvron2023llama}.

\noindent \textbf{Tool Call Ordering.} To guide the LLM, we encode a 
high-level debugging workflow into the prompt that mirrors how 
developers investigate crashes.
For global buffer overflows, the LLM is instructed to start from the 
buffer symbol reported in the ASan log, compute its size, 
and derive the buffer's end address for range validation. 
For heap buffer overflows, where ASan logs do not provide a symbol name, 
the LLM is guided to first apply buffer inference to locate candidate buffer 
references, print their addresses for validation, and 
fall back to scanning local variables and function arguments.

\noindent \textbf{Tool Error Tolerance.} Tools may return errors with unexpected inputs,
and without guidance the LLM may misinterpret these and repeat failing calls. 
To address this, we embed explicit error-handling instructions in the prompt. 
For example, in \autoref{list:bad-fix-example-2}, 
printing the address of \lstinline{CUR} returns a ``Symbol not found'' 
error since \lstinline{CUR} is a macro unavailable in the data view. 
LLM is instructed to invoke the macro resolution tool instead, 
avoiding repetitive and ineffective tool calls.

\subsection{Bug-Property-Guided Mitigation}
\label{ssec:design-repair}

Using the extracted context and validated \keyvar, \sys performs \contribone.
The LLM first instantiates the boundary condition in
\autoref{eq:bound-check}. A second step inserts this condition immediately
before the crash site and selects termination as the response. The output space is
therefore limited to adding a local guard rather than rewriting arbitrary
program logic.

We split this workflow into two agents. The first constructs the boundary
constraint from validated expressions. The second adapts the guard to the local
C or C++ context and corrects compilation errors. Neither agent is asked to
generate a permanent repair.

\noindent \textbf{Error-Message Driven Patch Correction.} Even when all variables are 
correctly identified, 
generating a compilable patch remains challenging 
due to compiler-specific restrictions across C and C++ projects.
For example, some compiler configurations disallow direct casting 
from \lstinline{char *} to \lstinline{void *}, 
and some buffer expressions are constant pointers that cannot be directly 
manipulated, causing an otherwise correct patch to fail compilation. 
To address this, \sys extracts the relevant portion of the compilation error message 
using regex patterns and provides it to the LLM alongside the failed patch, 
enabling targeted correction rather than blind regeneration.

\noindent \textbf{On-Demand Repetition.} Since the two agents are responsible
for distinct concerns, constraint generation and compilable guard insertion,
we categorize patch validation failures accordingly to avoid unnecessary repetition. 
Compilation failures are attributed to the patch-insertion agent, so only the second step 
is repeated. Crashes that persist after applying the patch indicate an 
incorrect boundary condition from the first agent, requiring the full workflow to be rerun. 
In our evaluation, 89\% of errors are compilation failures, so full reruns are rare.

\section{Implementation}
\label{sec:implement}

We implement \sys in approximately 5,500 lines of Python code. The current 
implementation accepts a reproduction environment and a PoC file that crash 
the program as input, and outputs both the results of key variable analysis and a candidate patch. 
\sys is designed to scale to diverse platforms. 
The current implementation leverages the ARVO format as input~\citep{mei2024arvo}.

\noindent \textbf{Step-By-Step Validation.} \sys validates each intermediate
output before proceeding. The Context Extractor first resolves
\lstinline{ptr}, then infers \lstinline{buffer_begin} and
\lstinline{buffer_end} from the validated pointer. Only after these expressions
pass their checks does \sys generate and insert the guard. If a step fails,
\sys retries that step or halts when no progress is made. This design avoids
building later LLM reasoning on rejected intermediate results and reduces the
risk of a plausible patch that is nevertheless an invalid mitigation.

\noindent \textbf{Debugging-Friendly Binary Preparation.} \sys relies on precise debugging information such as pointer types and memory addresses. Since the ARVO benchmark~\citep{mei2024arvo} and OSS-Fuzz~\citep{ossfuzz} enables aggressive compiler optimizations by default, \ie stripping debugging symbols, \sys intercepts the compilation process by replacing the compiler invocation with a custom Clang wrapper that enables full debugging information (\ie \lstinline{-O0 -g}).

\noindent \textbf{Environment-Independent Debugger Support.} \sys wraps GDB~\citep{gdb} to provide consistent debugging support across environments. Since bug reproduction environments span Ubuntu 16.04 with Clang 5.0.0 to Ubuntu 22.04 with Clang 15.0.0, the associated DWARF debug formats range from version 3 to 5. To ensure compatibility, \sys builds GDB 12.1~\citep{gdb_dwarf} from source, the minimum version supporting all DWARF formats used in our evaluation.

\noindent \textbf{Out-of-Box Mitigation Generation.} Prior systems typically co-locate the mitigation agent and the bug reproduction environment within the same container, complicating resource scheduling and leading to redundant dependency installations. \sys instead runs the mitigation framework on the host machine and interfaces with the reproduction environment via Docker, cleanly decoupling mitigation logic from bug reproduction and improving deployment flexibility.

\section{Evaluation}
\label{sec:evaluation}

\subsection{Setup}
\label{ssec:eval-setup}

\textbf{Hardware Environment.} All evaluations are conducted on a 
Desktop with Intel i7-13700 (24 threads, 5.20 GHz), 
64GB memory and 2TB SSD storage. 

\noindent \textbf{Large Language Model.} 
We employed Gemini-2.5-flash (June 2025) 
as the default LLM for all AVR agents~\citep{comanici2025gemini}.
Alternative models, including Gemini-2.5-pro and Gemini-2.5-flash-lite,
were tested but frequently failed due to internal errors 
(\eg “No generation chunks” and “internal server error” messages).
Consequently, Gemini-2.5-flash was selected for all experiments.

\noindent \textbf{Baseline AVR.} 
We compare \sys against state-of-the-art AVR agents PatchAgent~\citep{PatchAgent} and San2Patch~\citep{kim2025logs}. 
We exclude the agents whose source code is not available at the time of writing~\citep{zhang2024fixing,liu2025agent}. 
We also exclude constraint-based repairing tools~\citep{zhang2022program,shariffdeen2021concolic,gao2021beyond} as they follow a distinct system input: the accurate fault location, while \sys work 
without these results.

\noindent \textbf{Benchmark.} We utilize the ARVO~\citep{mei2024arvo} benchmark, 
comprised of over 5,000 real-world vulnerabilities found by OSS-Fuzz. 
Specifically, we choose \numarvocompare bugs following 
the recommended practice~\citep{zhang2024fixing}, specifically, 
the following three criteria: 
\begin{enumerate}
  \item \textbf{Reproducibility}: The bug must be reliable reproduced within
    15 minutes~\footnote{We change optimization from \lstinline{-O1} to 
    \lstinline{-O0 -g} for fully symbolized binary, 
    the modified compilation process encountered ASan errors.}.
  \item \textbf{Bug Type}: The bug represents a spatial memory corruption. 
  \item \textbf{Environment Compatibility}: The bug is compatible with \patchagent
  and \sanpatch's environment~\footnote{\patchagent and \sanpatch 
  do not support bugs that relies on Ubuntu 18 or before, 
  as they run a docker inside docker container. \sys supports all those bugs.}.
\end{enumerate}

\noindent \textbf{Metric.} 
For comparability with prior AVR studies, we retain the \textit{plausible patch}
metric~\citep{PatchAgent,zhang2024fixing}: a patch is plausible if replaying
the PoC no longer produces the sanitizer-reported error. In our framing, this is
evidence of mitigation for the observed execution, not proof of a correct
permanent repair or a generally safe mitigation. Because ARVO lacks
comprehensive functional tests, we separately audit the plausible patches in
\autoref{ssec:eval-study-manual}.

\noindent \textbf{Attempts, Repetition, and Validation.} For each agent, we evaluate two variants: $AgentName_1$, which runs a single attempt per trial, and $AgentName_5$, which runs five attempts per trial (following San2Patch~\citep{kim2025logs}). For $AgentName_1$, we run five independent trials and report the average to reduce the impact of randomness. In all settings, each attempt is allowed at most three patch validation calls~\citep{PatchAgent}.

\subsection{RQ1: How efficiently does \sys generate plausible patches?}
\label{ssec:eval-rq1}

\begin{figure*}[ht]
	\centering
	\includegraphics[width=\linewidth]{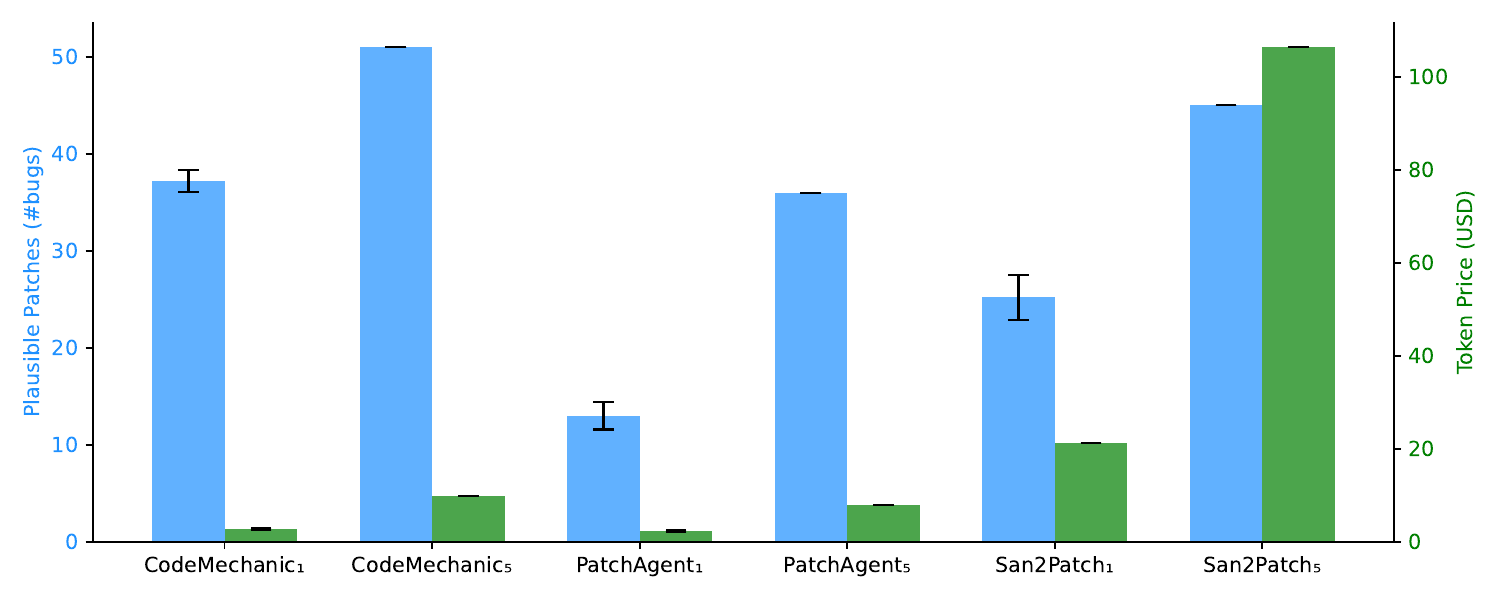}
\caption{Number of plausible patches and total token cost
  for agents on all \numarvocompare bugs. \sysfast, \patchfast and \sanfast
  are repeated for five times. 
  \greensquare\ stands for the cost (lower is better) while \bluesquare\  
  means the number of plausible patches (higher is better).}
	\label{fig:eval-rq1-num}
\end{figure*}

\autoref{fig:eval-rq1-num} compares the number of plausible patches and their
token cost across agents.

\textit{Comparative Plausible-Patch Count}. 
We compare \sys with other agents on the ARVO benchmark, as shown in~\autoref{fig:eval-rq1-num}. 
Overall, \sys demonstrates superior performance under limited numbers of attempts. In a single run, 
\sys achieves an average of 37.2 plausible patches across five runs, which is nearly three 
times that of \patchfast and 47.6\% higher than \sanfast.
When configured with more attempts, \sysfull maintains the leading performance,
outperforming \patchfull and \sanfull by 41.7\% and 13.3\%, respectively.

\textit{Cost Efficiency.} \sys also demonstrates strong cost efficiency. Under the single-attempt configuration, \sysfast incurs a total cost of \$2.75 to process all \numarvocompare bugs, compared to \$2.40 for \patchfast and \$21.3 for \sanfast. Although \sanfast produces nearly twice as many plausible patches as \patchfast, it does so at nearly ten times the cost. In contrast, \sysfast achieves a more favorable trade-off: at only 13\% of the cost of \sanfast, it generates 47.6\% more plausible patches. Under the full five-attempt configuration, \sysfull costs \$9.90 and produces 51 plausible patches, whereas \sanfull costs \$106.39, 9.7$\times$ more, while producing 10\% fewer plausible patches. \patchfull remains the most economical option at \$7.92, but produces the fewest plausible patches. By spending only \$2 more than \patchfull, \sysfull produces 15 additional plausible patches. This cost efficiency is useful for rapidly generating a provisional mitigation. RQ4 separately audits its safety and semantic equivalence.

\begin{figure*}[t]
	\centering
	\includegraphics[width=0.7\linewidth]{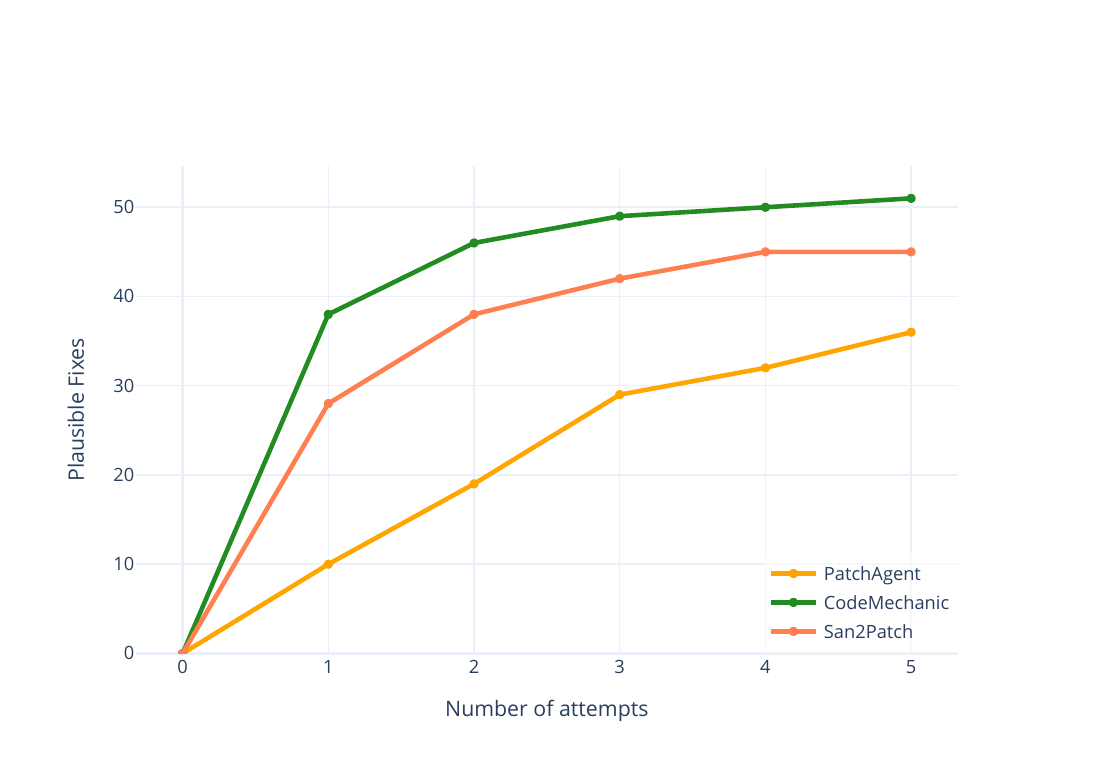}
\caption{Total number of plausible patches generated at the n-th attempt. 
  The grey dot lines stands for the plausible patches achieved at the first and fifth 
  \sysfull attempt.}
	\label{fig:eval-rq1-attempts}
\end{figure*}

\textit{Plausible Patches per Attempt}. 
To further evaluate the benefit of multiple attempts, we examine the cumulative number of plausible patches across attempts, as shown in~\autoref{fig:eval-rq1-attempts}. \sysfull demonstrates consistently higher efficiency: its first attempt alone outperforms \patchfull across all five attempts, highlighting the effectiveness of each individual patch generation attempt. Beyond fixed configurations, \sys supports adjustable attempt counts, allowing users to trade cost for performance by increasing the number of attempts beyond the five-attempt setting evaluated here.

\textit{Overlap Analysis.} 
We also explore the overlap of plausible patches generated by three agents. %
Overall, different agents produce plausible patches for distinct bugs.
Together, the agents produce plausible patches for 81 distinct bugs, but only
8 bugs receive plausible patches from every agent.
This complementarity supports a two-stage workflow. The first stage uses \sys
to obtain a constrained mitigation. The second passes cases without a candidate
patch to an open-ended repair agent.
Relative to \sys alone, this combination produces
plausible patches for 41\% more bugs at an additional cost of \$0.93.

\begin{formal}
  \textbf{Takeaway:} 
  \sysfast generates 47.6\% more plausible patches than the second-best
  agent at approximately 10\% of its cost.
\end{formal}

\subsection{RQ2: What types of bugs does \sys mitigate?}
\label{ssec:eval-bug-type}

\begin{table}[t!]
  \centering
\resizebox{0.9\textwidth}{!} {
\begin{tabular}{ll|rrrrrr|r}
\toprule
\multicolumn{2}{l|}{\multirow{2}{*}{Agent}} & \multicolumn{2}{c}{\sys} & \multicolumn{2}{c}{\patchagent} & \multicolumn{2}{c|}{\sanpatch} & \multicolumn{1}{c}{\multirow{2}{*}{Total Bugs}} \\
\multicolumn{2}{l|}{}                       & fast            & full           & fast           & full          & fast           & full          & \multicolumn{1}{c}{}                            \\ \midrule
\multirow{3}{*}{Bug Type}      & Stack      & 1.4             & 2              & 1.8            & 5             & 2              & 4             & 8                                               \\
                               & Heap       & 27.2            & 40             & 9.8            & 25            & 18.8           & 35            & 81                                              \\
                               & Global     & 8.6             & 9              & 1.4            & 6             & 4.4            & 6             & 12                                              \\ \midrule
\multicolumn{2}{l|}{Total}                  & 37.2            & 51             & 13             & 36            & 25.2           & 45            & 101                                             \\ \midrule
\end{tabular}
}
  \caption{
    Average plausible patches generated by \patchagent, \sanpatch and \sys. 
    fast means one attempt and full means five attempts.
  }
  \label{tab:bug-class-plausible}
\end{table}

We further examine the mitigation performance broken down by vulnerability category, as summarized in~\autoref{tab:bug-class-plausible}.

Stack-based buffer overflows remain the most challenging category for \sys. Among the eight stack-based cases, \sysfull produces at most two plausible patches, whereas \patchfull and \sanfull produce five and four, respectively. Notably, across multiple attempts, \patchfull and \sanfull tend to mitigate different stack bugs, while \sys consistently addresses the same ones, suggesting that \sys exhibits more deterministic behavior compared to the higher stochasticity of the other agents.

Heap-based buffer overflows favor \sys. \sysfast markedly outperforms \patchfast and \sanfast in the single-attempt setting, and \sysfull maintains this advantage with approximately 60\% and 14\% more plausible patches than \patchfull and \sanfull, respectively.

Global buffer overflows highlight the stability of \sys. \sysfast and \sysfull produce nearly identical results, 8.6 and 9 plausible patches, indicating robust and consistent performance on this class. In contrast, \patchfast produces only 1.4 plausible patches on average, improving to 6 with multiple attempts, while \sanfast averages 4.4 plausible patches and \sanfull produces 6.

\begin{formal}
  \textbf{Takeaway:} \sys achieves the most stable performance across heap and global buffer overflows, 
  while stack-based bugs remain challenging.
\end{formal}

\subsection{RQ3: At which stage does mitigation generation fail?}
\label{ssec:eval-failed-stage}

\begin{figure*}[t]
	\centering
	\includegraphics[width=0.9\linewidth]{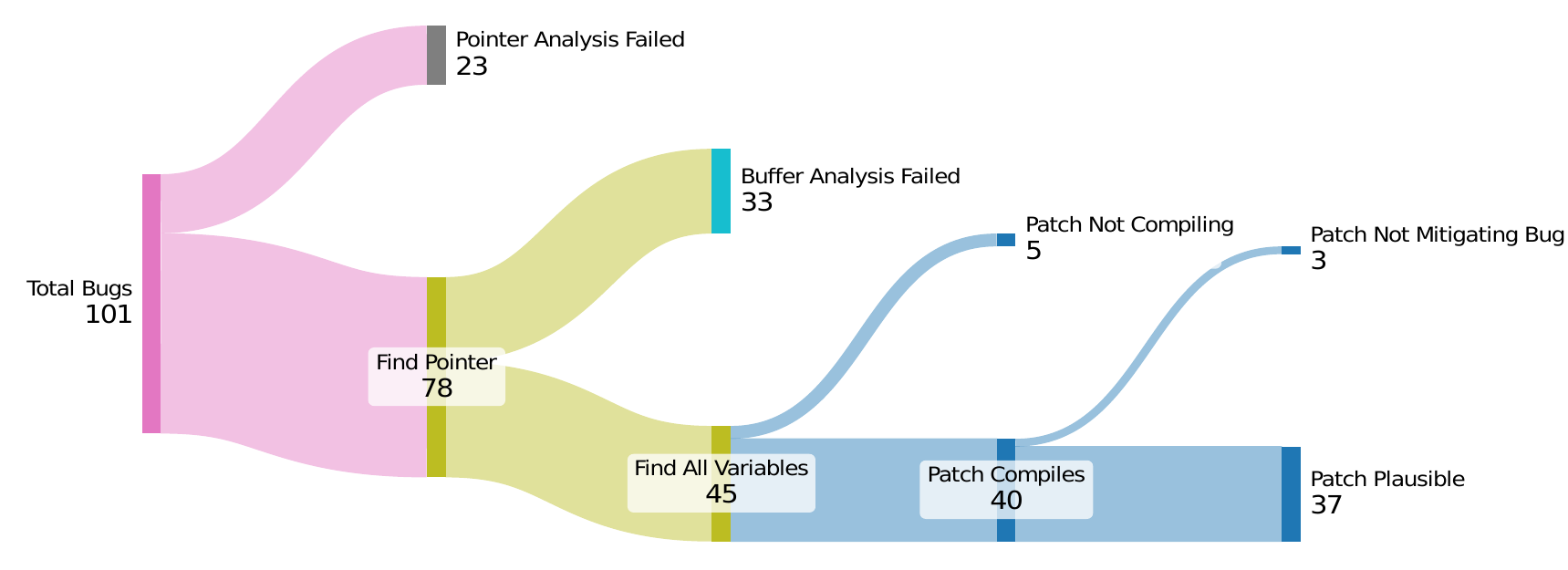}
	\caption{Pipeline stage at which \sysfast fails to generate a candidate patch.
  \patchagent and \sanpatch do not follow our workflow, thus cannot be analyzed.}
	\label{fig:eval-rq2-failed}
\end{figure*}

We further investigate the stage at which \sys fails in~\autoref{fig:eval-rq2-failed}.

\sys starts with its Context Extractor. It analyzes the pointer that 
violates the memory safety and infers the corresponding buffer range. 
In this phase, \sys finds the pointers for most of the bugs, and only 
fails for 23 out of \numarvocompare bugs. Buffer analyses are more challenging, 
even with the buffer usage oracles and debugging tools, 
around 42\% of buffer analyses failed. 
The prime reason is that buffer definitions usually occur in a different part of the 
program, far away from the place where pointer is dereferenced. 

After identifying all \keyvar, the patch generation becomes simpler. 
Given the 45 bugs for which \sys finds all \keyvar, 37 plausible patches are generated.
Thus, when the Context Extractor recovers all required expressions, the LLM
generates a plausible mitigation for 82\% of cases within three attempts. Most
remaining failures are compilation failures rather than failures to instantiate
the boundary condition.

\begin{formal}
\begin{minipage}{\linewidth}
\textbf{Takeaway:} Context extraction remains challenging
for spatial memory corruption. Given all \keyvar, \sys generates a plausible
mitigation in more than 82\% of cases.
\end{minipage}
\end{formal}

\subsection{RQ4: How many "plausible" patches are "correct"?}
\label{ssec:eval-study-manual}

We manually audit all plausible patches along two dimensions: whether they
provide a functionally acceptable mitigation for the reported access, and
whether they match the broader behavior of the developer-written repair. We classify a
patch as \textit{Semantically Equivalent (SE)} if it is functionally
identical to the developer-written repair, or \textit{SE with Redundant Checks (SERC)} if
it adds harmless safety checks. \textit{Correct Boundary (CB)} patches
mitigate the reported access but do not address other issues handled by the
developer-written repair. \textit{CB with Silent Error (CBSE)} patches prevent the
access but may silently misprocess malformed inputs. Finally,
\textit{Incorrect Patches (IP)} pass PoC-replay validation but break intended
functionality. In aggregate results, ``semantically equivalent'' includes SE and
SERC, while ``correct'' includes all non-IP categories.

\begin{table}[]
  \resizebox{\textwidth}{!} {
  \begin{tabular}{l|cccc}
  \toprule
  Status                        & \multicolumn{1}{c}{\begin{tabular}[c]{@{}c@{}}Inexplict Assumption \\ Awarenes\end{tabular}} & \multicolumn{1}{c}{\begin{tabular}[c]{@{}c@{}}Mitigate Different \\ Bugs\end{tabular}} & \multicolumn{1}{c}{\begin{tabular}[c]{@{}c@{}}Functionally \\ Correct\end{tabular}} & \multicolumn{1}{c}{\begin{tabular}[c]{@{}c@{}}Mitigate Current \\ Bug\end{tabular}} \\ \midrule
  Semantically Equivalent (SE)      & \cmark & \cmark & \cmark & \cmark                                                                                 \\
  SE w/ Redundant Checks (SERC) & \xmark & \cmark & \cmark & \cmark                                                                                 \\
  Correct Boundary (CB)         & \xmark & \xmark & \cmark & \cmark                                                                                 \\
  CB w/ Silent Error (CBSE)         & \xmark & \xmark & \halfcheckmark $^{*}$ & \cmark                                                                                 \\

  Incorrect Patch (IP)              & \xmark & \xmark & \xmark & \cmark                                                                                 \\ \midrule
  \end{tabular}
  }
  \caption{Manual analysis status. *: only guarantee functional correctness 
  with expected input.}
  \label{tab:manual-label}
\end{table}

\begin{figure*}[!t]
	\centering
	\includegraphics[width=\linewidth]{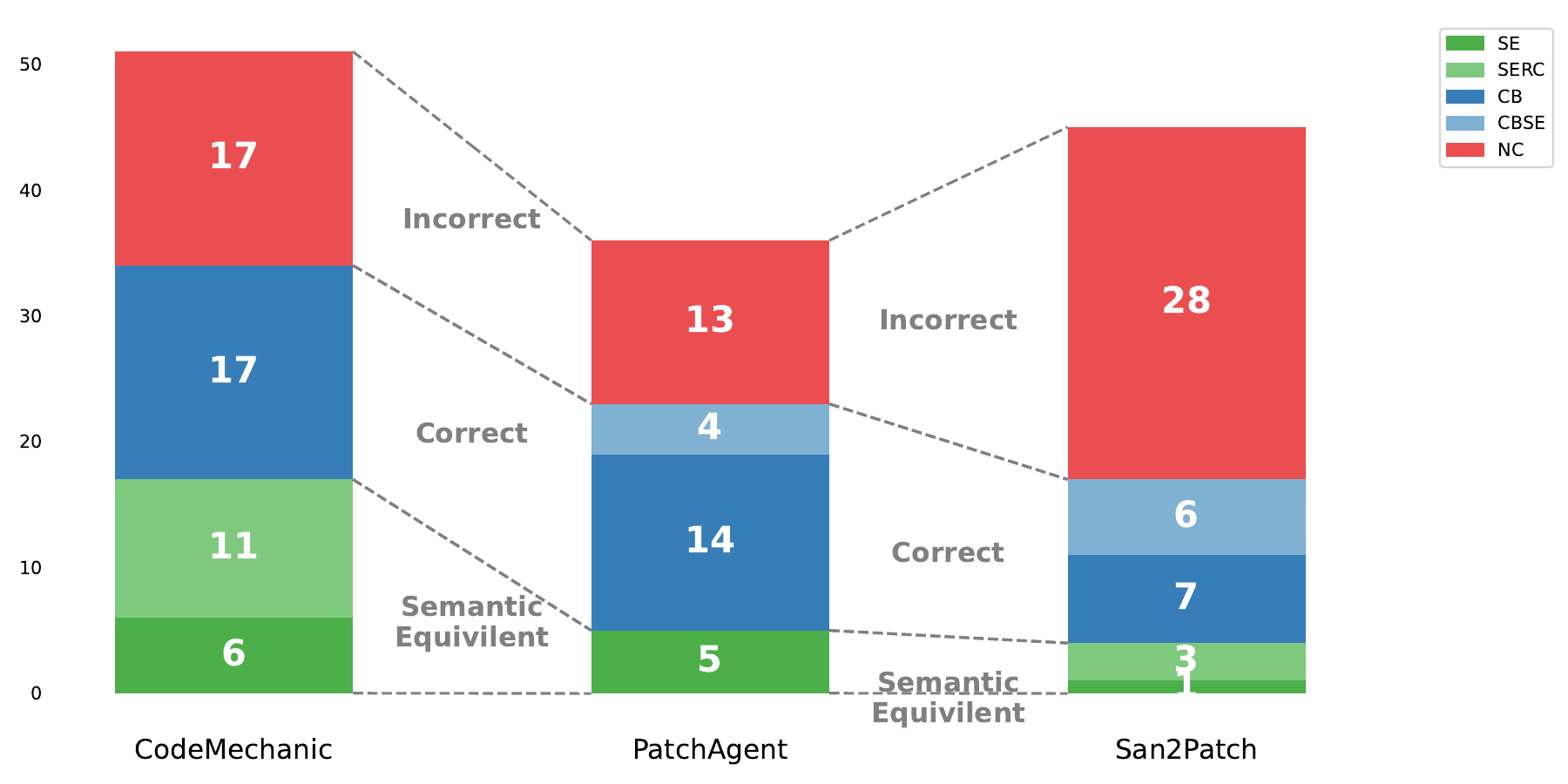}
	\caption{Manual audit results of 'plausible' patches generated by \sysfull, \patchfull and \sanfull.
  SE denotes Semantically Equivalent, SERC denotes SE with redundant checks, CB denotes Correct
  Boundary, CBSE denotes CB with Silent Error, and NC denotes Not Correct, corresponding to IP in~\autoref{tab:manual-label}.}
	\label{fig:eval-rq4-audit}
\end{figure*}

\autoref{fig:eval-rq4-audit} presents the audit results. Of the 51 plausible patches
from \sysfull, 34 are non-IP mitigations and 17 are semantically equivalent to
the developer-written repair, comprising 6 SE and 11 SERC patches. \patchfull
produces 23 non-IP patches, including 5 SE patches, while \sanfull
produces 17 non-IP patches, including 1 SE and 3 SERC patches. Thus, passing
PoC-replay validation alone substantially overestimates both mitigation safety and
equivalence to the developer-written repair.

The agents also fail differently. \sysfull consistently places a boundary guard
immediately before the dangerous access and terminates when the guard is
violated. This converts the detected memory corruption into an explicit
availability failure. In contrast, \patchfull and \sanfull sometimes silently
discard or truncate malformed input, producing four and six CBSE patches,
respectively. Their open-ended patches can also modify locations unrelated to
the vulnerable access. The local guard of \sys limits that freedom, although an
incorrectly inferred boundary can still terminate valid executions, as examined
in RQ5.

\begin{formal}
  \textbf{Takeaway:} Passing PoC-replay validation does not imply a safe mitigation or a
  permanent repair. \sys produces 34 non-IP mitigations, including 17 that are
  semantically equivalent to developer-written repairs.
\end{formal}

\subsection{RQ5: Why do plausible patches generated by \sys break functionality?}
\label{ssec:eval-failure-casestudy}

We further analyze plausible patches that break program
functionality.
Two primary failures were identified:

\begin{figure*}[t!] 
  \begin{lstlisting}[style=cpp]
// OSS-VUL-64110: Finer-Grained Control
struct gre1_header : gre_basic_header {
  uint16_t payloadLength;
  uint16_t callID;
};
// ASan boundary is [gh_p, gh_p + sizeof(gre1_header))
// CodeMechanic boundary is [p_len, p_len + sizeof(uint16_t))
char *p_len = &gh_p->payloadLength;

// OSS-VUL-42244: Over-Restricted Boundaries
// ASan boundary is [s, s + s_len)
// CodeMechanic boundary is [s, s + i], i < s_len
if (&s[i] < s || &s[i] >= (char *)s + i) {
  exit(0);
}

\end{lstlisting}
\caption{\sys failure 1: Over-Restricted Boundaries.}
\label{list:failure-case-1}
\end{figure*}

\noindent  \textbf{Over-Restricted Boundaries.}
\sys currently accepts a generated boundary when it is a subset of the
ASan-detected allocation.
For example, in \autoref{list:failure-case-1}, ASan reports the allocation
\lstinline{[gh_p, gh_p + sizeof(gre1_header))}, while \sys selects the smaller
range \lstinline{[&gh_p->payloadLength, &gh_p->payloadLength + sizeof(uint16_t))}.
A legitimate subobject may require such a finer-grained range. However, the same
acceptance rule can admit a nonexistent range such as \lstinline{[s, s + i)}.
The resulting guard triggers too often, converting valid executions into denial
of service. This over-restrictive guard can cause denial of service, but it still
prevents the guarded memory corruption while developers investigate and prepare
a permanent repair.

\noindent \textbf{Hardcoded Variables.}
\sys integrates runtime debugging information to infer actual
boundaries. However, it may unintentionally incorporate concrete
runtime values during this process.
Specifically, \autoref{list:failure-case-2} demonstrates a case where the
analysis identifies the symbolic boundary
\lstinline{[dat->chain, dat->chain + dat->size)}, but the generated condition
also embeds an absolute address from the observed execution.
Although \sys explicitly instructs the LLM to avoid using absolute values,
the model occasionally produces patches containing absolute addresses or
runtime-specific size values. Such a guard may pass PoC-replay validation but fail
on another execution, leaving the bug exploitable.

\begin{figure*}[t!] 
  \begin{lstlisting}[style=cpp]
// OSS-VUL-63463 Fix
// CodeMechanic use constant address instead of variable
if (&dat->chain[dat->byte] < dat->chain || 
(char*)0x61d00000d1f0 > (char *)dat->chain + dat->size) {
  exit(0);
}
\end{lstlisting}
\caption{\sys failure 2: Hardcoded Variables.}
\label{list:failure-case-2}
\end{figure*}

These failures expose two different deployment tradeoffs. Matching only the
complete ASan allocation would reject valid subobject bounds, while rejecting
every constant would also reject legitimate expressions such as
\lstinline{sizeof(buf)}. A guard that embeds constants requires human review
because it may not work for other inputs.

\begin{formal}
  \textbf{Takeaway:} 
  Invalid patches primarily arise from over-restricted ranges, which trade
  availability for security, or runtime-specific constants, which may
  leave the bug exploitable and therefore require human review.
\end{formal}

\section{Related Works}
\label{sec:related-work}

\noindent \textbf{Bug Discovery.}
A prerequisite for AVR is the ability to identify bugs. Over the past 
decades, numerous bug-finding techniques 
have been developed, including static analysis~\citep{codeql,yamaguchi2014modeling,yamaguchi2013chucky} 
and greybox fuzzing~\citep{afl,fioraldi2020afl++,libfuzzer}. Greybox 
fuzzing, in particular, has gained prominence as it minimizes the 
need for human expertise while automating bug discovery. As a result, 
it has been extensively studied in both academia~\citep{bohme2016coverage,zheng2023fishfuzz,zheng2025mendelfuzz,she2022effective} and 
industry~\citep{clusterfuzz}.
When combined with sanitizers~\citep{serebryany2012addresssanitizer}, 
modern fuzzing infrastructures such as ClusterFuzz~\citep{clusterfuzz} 
automatically generate bug reports containing minimized bug-inducing 
inputs and reproducible environments, including PoC 
and automated build scripts. These reproducible environments 
facilitate developer analysis and manual repair, while also providing strong
foundations for bug-property-guided mitigation.
To align with real-world practices, \sys adopts bug reports and their 
corresponding environments from OSS-Fuzz~\citep{ossfuzz}, one of the 
most widely deployed industrial fuzzing platforms.

\noindent \textbf{LLM-Based Automated Vulnerability Repair.}
PatchAgent and San2Patch~\citep{PatchAgent,kim2025logs} use sanitizer logs for
localization and root-cause analysis, while VulDebugger~\citep{liu2025agent}
incorporates dynamic debugging. These systems pursue unrestricted LLM-generated
repairs intended to match developer-written repairs and therefore leave substantial freedom to the LLM. Their final oracle
is commonly whether the patched program still crashes on the PoC. \sys takes a
different approach. It uses ASan's debugging API~\citep{asan_debugger} to
check intermediate boundary expressions and limits generation to a local guard.
This restriction does not attempt a complete repair. In exchange, it
provides a more predictable temporary mitigation.

\section{Threats to Validity}
\label{sec:discussion}

Our results measure bug-property-guided mitigation on a selected benchmark and do not
establish that every generated patch is safe for automatic production
deployment. We discuss the main limitations below.

\subsection{Internal Threats.}
\label{ssec:discuss-interal}

\noindent \textbf{Limited Validation.} \sys validates buffer boundaries using sanitizer metadata, which confirms that a buffer range is the nearest valid region to the pointer but does not guarantee it is the \emph{intended} buffer. For example, a pointer intending to access \lstinline{[addr1, addr1+0x100)} but positioned at \lstinline{&addr1[0x1FF]} could be misclassified as belonging to an adjacent buffer \lstinline{[addr1+0x200, addr1+0x300)}. In practice, however, most memory corruptions involve small overflows of only a few bytes, and we configure ASan with a conservative 2048-byte redzone to further reduce such misclassifications. %

\noindent \textbf{Overfitting the Training Set.} Since LLMs are trained on public code repositories, their training data may include previously disclosed vulnerabilities and patches, risking overfitting. To mitigate this, \sys decomposes patch generation into three independent steps ( pointer and buffer analysis, constraint generation, and patch insertion) without exposing any project- or program-specific information. In particular, the constraint generation step provides only variable names and constraint requirements, substantially limiting the LLM's exposure to identifiable context.

\subsection{External Threats.}
\label{ssec:discuss-external}

\noindent \textbf{Limited Bug Categories.} \sys is currently focused on spatial memory corruptions, motivated by their security severity and prevalence: 25 out of 31 in-the-wild zero-day exploits in 2025 were memory corruptions~\citep{0dayinthewild}, and 65.8\% of ASan errors in the OSS-Fuzz benchmark are buffer overflows~\citep{mei2024arvo}. While \sys could in principle be extended to temporal memory corruptions, this is currently hindered by the lack of reliable detection oracles, such as determining whether a buffer has been freed.

\noindent \textbf{Defensive Termination.} The current patch policy
intentionally terminates execution when a boundary violation is detected. For
exploitable memory corruptions, this mitigation strategy trades availability for security by converting a potential remote code execution into a denial of service. Such termination behavior is also used in production security practice, where intentional termination is preferred to continued execution in a compromised state~\citep{chrome_checklist}. The generated patch therefore provides an immediate safeguard while giving maintainers more time to investigate the root cause and develop an application-specific repair.

\noindent \textbf{Target-Specific Human Debugging Knowledge.} \sys embeds high-level debugging knowledge into the prompt to guide tool invocation and error handling, with a single prompt designed per bug type. To limit overfitting, we use abstract instructions rather than program-specific guidance, and develop prompts using only ten bugs per category. Evaluation results suggest this knowledge transfers well to the broader benchmark of \numarvocompare bugs.

\noindent \textbf{PoC-Based Mitigation Oracle.} The widely used
\emph{plausible} metric~\citep{PatchAgent,zhang2024fixing} establishes only that
the sanitizer-reported error disappears when the PoC is replayed. It does not
prove that the mitigation blocks other exploit inputs or that valid behavior is
preserved. Unit tests used by recent systems~\citep{PatchAgent,kim2025logs} may
also miss the modified region. For example, the functionality-breaking change
in \autoref{list:bad-fix-example} can pass tests that do not exercise the fuzz
driver. We therefore audit all plausible patches in
\autoref{ssec:eval-study-manual}. 

\section{Conclusion}
\label{sec:conclusion}

We present \sys, a bug-property-guided system for constrained mitigation of spatial
memory corruption. Rather than asking an LLM to generate a permanent
repair, \sys reconstructs a boundary property, validates its key
expressions, and inserts a local guard before the dangerous access.
This design intentionally trades availability for security, providing a
temporary mitigation while developers prepare a permanent repair.
On 101 ARVO bugs, \sys generates more plausible patches at substantially
lower cost than \patchagent and \sanfull, and it produces 240\% and 325\% more
semantically equivalent patches, respectively.
We release the prototype to support open science and further research.

\subsubsection*{Broader Impact Statement}
Our research does not involve the discovery of new vulnerabilities. Instead, this work proposes a mitigation-generation framework that reduces the period during which a disclosed vulnerability remains exploitable, thereby supporting defensive techniques that strengthen software security.

\bibliography{paper}

@misc{afl,
	author = "Zalewski, Michal",
	title = "american fuzzy lop",
	howpublished = "\url{https://lcamtuf.coredump.cx/afl/}",
	year={2013}
}

@inproceedings{fioraldi2020afl++,
  title={AFL++ combining incremental steps of fuzzing research},
  author={Fioraldi, Andrea and Maier, Dominik and Ei{\ss}feldt, Heiko and Heuse, Marc},
  booktitle={Proceedings of the 14th USENIX Conference on Offensive Technologies},
  pages={10--10},
  year={2020}
}

@misc{libfuzzer,
	author = "libfuzzer",
	title = "libfuzzer",
	howpublished = "\url{https://llvm.org/docs/LibFuzzer.html}",
	year={2023}
}

@inproceedings{zheng2023fishfuzz,
  title={$\{$FISHFUZZ$\}$: Catch deeper bugs by throwing larger nets},
  author={Zheng, Han and Zhang, Jiayuan and Huang, Yuhang and Ren, Zezhong and Wang, He and Cao, Chunjie and Zhang, Yuqing and Toffalini, Flavio and Payer, Mathias},
  booktitle={32nd USENIX Security Symposium (USENIX Security 23)},
  pages={1343--1360},
  year={2023}
}

@misc{syzbot_num_vm,
	author = {Aleksandr Nogikh},
	title = "Syzbot: 7 years of continuous kernel fuzzing",
	howpublished = "\url{https://lpc.events/event/17/contributions/1521/attachments/1272/2698/LPC23_Syzbot_years_of_continuous_kernel_fuzzing.pdf}",
	year={2023}
}

@misc{syzbot,
	author = {Google},
	title = "syzbot",
	howpublished = "\url{https://syzkaller.appspot.com/upstream}",
	year={2024}
}

@article{tufano2019empirical,
  title={An empirical study on learning bug-fixing patches in the wild via neural machine translation},
  author={Tufano, Michele and Watson, Cody and Bavota, Gabriele and Penta, Massimiliano Di and White, Martin and Poshyvanyk, Denys},
  journal={ACM Transactions on Software Engineering and Methodology (TOSEM)},
  volume={28},
  number={4},
  pages={1--29},
  year={2019},
  publisher={ACM New York, NY, USA}
}

@article{chen2019sequencer,
  title={Sequencer: Sequence-to-sequence learning for end-to-end program repair},
  author={Chen, Zimin and Kommrusch, Steve and Tufano, Michele and Pouchet, Louis-No{\"e}l and Poshyvanyk, Denys and Monperrus, Martin},
  journal={IEEE Transactions on Software Engineering},
  volume={47},
  number={9},
  pages={1943--1959},
  year={2019},
  publisher={IEEE}
}

@article{feng2020codebert,
  title={Codebert: A pre-trained model for programming and natural languages},
  author={Feng, Zhangyin and Guo, Daya and Tang, Duyu and Duan, Nan and Feng, Xiaocheng and Gong, Ming and Shou, Linjun and Qin, Bing and Liu, Ting and Jiang, Daxin and others},
  journal={arXiv preprint arXiv:2002.08155},
  year={2020}
}

@inproceedings{huang2025template,
  title={Template-guided program repair in the era of large language models},
  author={Huang, Kai and Zhang, Jian and Meng, Xiangxin and Liu, Yang},
  year={2025},
  organization={ICSE}
}

@inproceedings{xia2022less,
  title={Less training, more repairing please: revisiting automated program repair via zero-shot learning},
  author={Xia, Chunqiu Steven and Zhang, Lingming},
  booktitle={Proceedings of the 30th ACM Joint European Software Engineering Conference and Symposium on the Foundations of Software Engineering},
  pages={959--971},
  year={2022}
}

@inproceedings{xia2024automated,
  title={Automated program repair via conversation: Fixing 162 out of 337 bugs for \$0.42 each using ChatGPT},
  author={Xia, Chunqiu Steven and Zhang, Lingming},
  booktitle={Proceedings of the 33rd ACM SIGSOFT International Symposium on Software Testing and Analysis},
  pages={819--831},
  year={2024}
}

@article{zhang2024fixing,
  title={Fixing Security Vulnerabilities with AI in OSS-Fuzz},
  author={Zhang, Yuntong and Wang, Jiawei and Berzin, Dominic and Mirchev, Martin and Liu, Dongge and Arya, Abhishek and Chang, Oliver and Roychoudhury, Abhik},
  journal={arXiv preprint arXiv:2411.03346},
  year={2024}
}

@inproceedings{huang2023empirical,
  title={An empirical study on fine-tuning large language models of code for automated program repair},
  author={Huang, Kai and Meng, Xiangxin and Zhang, Jian and Liu, Yang and Wang, Wenjie and Li, Shuhao and Zhang, Yuqing},
  booktitle={2023 38th IEEE/ACM International Conference on Automated Software Engineering (ASE)},
  pages={1162--1174},
  year={2023},
  organization={IEEE}
}

@misc{memory_corruption_exploitable, 
  title = {The More You Know, The More You Know You Don’t Know - A Year in Review of 0-days Used In-the-Wild in 2021},
  url = {https://googleprojectzero.blogspot.com/2022/04/the-more-you-know-more-you-know-you.html},
  author = {Maddie Stone},
  year = {2022}
}

@article{mei2024arvo,
  title={ARVO: Atlas of Reproducible Vulnerabilities for Open Source Software},
  author={Mei, Xiang and Singaria, Pulkit Singh and Del Castillo, Jordi and Xi, Haoran and Bao, Tiffany and Wang, Ruoyu and Shoshitaishvili, Yan and Doup{\'e}, Adam and Pearce, Hammond and Dolan-Gavitt, Brendan and others},
  journal={arXiv preprint arXiv:2408.02153},
  year={2024}
}

@misc{asan_debugger,
	author = {Google},
	title = "AddressSanitizerAndDebugger",
	howpublished = "\url{https://github.com/google/sanitizers/wiki/AddressSanitizerAndDebugger}",
	year={2016}
}

@inproceedings{serebryany2012addresssanitizer,
  title={$\{$AddressSanitizer$\}$: A fast address sanity checker},
  author={Serebryany, Konstantin and Bruening, Derek and Potapenko, Alexander and Vyukov, Dmitriy},
  booktitle={2012 USENIX annual technical conference (USENIX ATC 12)},
  pages={309--318},
  year={2012}
}

@article{PatchAgent,
  title     = {PatchAgent: A Practical Program Repair Agent Mimicking Human Expertise},
  author    = {Yu, Zheng and Guo, Ziyi and Wu, Yuhang and Yu, Jiahao and 
               Xu, Meng and Mu, Dongliang and Chen, Yan and Xing, Xinyu},
  booktitle = {34rd USENIX Security Symposium (USENIX Security 25)},
  year      = {2025}
}

@misc{gdb,
	author = {The GNU Project},
	title = "GDB: The GNU Project Debugger",
	howpublished = "\url{https://sourceware.org/gdb/}",
	year={2025}
}

@misc{gdb_dwarf,
	author = {The GNU Project},
	title = "GDB 12 DWARF-5 support",
	howpublished = "\url{https://gcc.gnu.org/onlinedocs/gcc/Debugging-Options.html/}",
	year={2025}
}

@article{liu2025agent,
  title={Agent That Debugs: Dynamic State-Guided Vulnerability Repair},
  author={Liu, Zhengyao and Ma, Yunlong and Xu, Jingxuan and Ai, Junchen and Gao, Xiang and Sun, Hailong and Roychoudhury, Abhik},
  journal={arXiv preprint arXiv:2504.07634},
  year={2025}
}

@misc{ossfuzz,
	author = "google",
	title = "ossfuzz",
	howpublished = "\url{https://google.github.io/ossfuzz/}",
	year={2023}
}

@misc{chrome_checklist,
  author = "Chrome",
  title = "Chrome Security Checklist",
  howpublished = "\url{https://chromium.googlesource.com/chromium/src/+/main/docs/security/checklist.md}",
  year={2025}
}

@article{comanici2025gemini,
  title={Gemini 2.5: Pushing the frontier with advanced reasoning, multimodality, long context, and next generation agentic capabilities},
  author={Comanici, Gheorghe and Bieber, Eric and Schaekermann, Mike and Pasupat, Ice and Sachdeva, Noveen and Dhillon, Inderjit and Blistein, Marcel and Ram, Ori and Zhang, Dan and Rosen, Evan and others},
  journal={arXiv preprint arXiv:2507.06261},
  year={2025}
}

@article{liu2024deepseek,
  title={Deepseek-v3 technical report},
  author={Liu, Aixin and Feng, Bei and Xue, Bing and Wang, Bingxuan and Wu, Bochao and Lu, Chengda and Zhao, Chenggang and Deng, Chengqi and Zhang, Chenyu and Ruan, Chong and others},
  journal={arXiv preprint arXiv:2412.19437},
  year={2024}
}

@misc{chatgpt,
  author = "OpenAI",
  title = "ChatGPT",
  howpublished = "\url{https://chatgpt.com/}",
  year={2025}
}

@inproceedings{qi2015analysis,
  title={An analysis of patch plausibility and correctness for generate-and-validate patch generation systems},
  author={Qi, Zichao and Long, Fan and Achour, Sara and Rinard, Martin},
  booktitle={Proceedings of the 2015 international symposium on software testing and analysis},
  pages={24--36},
  year={2015}
}

@article{hack_stack1996,
  title={Hacking the Stack for Fun and Profit},
  author={Aleph One},
  journal={Phrack Magazine},
  year={1996}
}

@misc{memory_safety_chrome,
  title = {Memory safety in Chromium},
  url = {https://www.chromium.org/Home/chromium-security/memory-safety/},
  author = {Chromium Security Team},
  year = {2025}
}

@inproceedings{carlini2015control,
  title={$\{$Control-Flow$\}$ bending: On the effectiveness of $\{$Control-Flow$\}$ integrity},
  author={Carlini, Nicholas and Barresi, Antonio and Payer, Mathias and Wagner, David and Gross, Thomas R},
  booktitle={24th USENIX Security Symposium (USENIX Security 15)},
  pages={161--176},
  year={2015}
}

@inproceedings{wahbe1993efficient,
  title={Efficient software-based fault isolation},
  author={Wahbe, Robert and Lucco, Steven and Anderson, Thomas E and Graham, Susan L},
  booktitle={Proceedings of the fourteenth ACM symposium on Operating systems principles},
  pages={203--216},
  year={1993}
}

@misc{harden_allocator,
  title = {ScudoHardenedAllocator},
  url = {https://llvm.org/docs/ScudoHardenedAllocator.html},
  author = {LLVM Team},
  year = {2025}
}

@misc{too_many_bugs_fix,
	author = "Angelos Keromytis",
	title = "Recommendations from the Workshop on Open-source Software Security Initiative",
	howpublished = "\url{https://bpb-us-e1.wpmucdn.com/sites.gatech.edu/dist/a/2878/files/2022/10/OSSI-Final-Report.pdf}",
	year={2022}
}

@article{touvron2023llama,
  title={Llama: Open and efficient foundation language models},
  author={Touvron, Hugo and Lavril, Thibaut and Izacard, Gautier and Martinet, Xavier and Lachaux, Marie-Anne and Lacroix, Timoth{\'e}e and Rozi{\`e}re, Baptiste and Goyal, Naman and Hambro, Eric and Azhar, Faisal and others},
  journal={arXiv preprint arXiv:2302.13971},
  year={2023}
}

@misc{codeql,
	author = "GitHub",
	title = "CodeQL: the libraries and queries that power security researchers around the world",
	howpublished = "\url{https://codeql.github.com/}",
	year={2024}
}

@inproceedings{yamaguchi2014modeling,
  title={Modeling and discovering vulnerabilities with code property graphs},
  author={Yamaguchi, Fabian and Golde, Nico and Arp, Daniel and Rieck, Konrad},
  booktitle={2014 IEEE symposium on security and privacy},
  pages={590--604},
  year={2014},
  organization={IEEE}
}

@inproceedings{yamaguchi2013chucky,
  title={Chucky: Exposing missing checks in source code for vulnerability discovery},
  author={Yamaguchi, Fabian and Wressnegger, Christian and Gascon, Hugo and Rieck, Konrad},
  booktitle={Proceedings of the 2013 ACM SIGSAC conference on Computer \& communications security},
  pages={499--510},
  year={2013}
}

@misc{clusterfuzz,
	author = "google",
	title = "ClusterFuzz",
	howpublished = "\url{https://google.github.io/clusterfuzz/}",
	year={2023}
}

@inproceedings{bohme2016coverage,
  title={Coverage-based greybox fuzzing as markov chain},
  author={B{\"o}hme, Marcel and Pham, Van-Thuan and Roychoudhury, Abhik},
  booktitle={Proceedings of the 2016 ACM SIGSAC Conference on Computer and Communications Security},
  pages={1032--1043},
  year={2016}
}

@article{zheng2025mendelfuzz,
  title={MendelFuzz: The Return of the Deterministic Stage},
  author={Zheng, Han and Toffalini, Flavio and B{\"o}hme, Marcel and Payer, Mathias},
  journal={Proceedings of the ACM on Software Engineering},
  volume={2},
  number={FSE},
  pages={44--64},
  year={2025},
  publisher={ACM New York, NY, USA}
}

@inproceedings{she2022effective,
  title={Effective seed scheduling for fuzzing with graph centrality analysis},
  author={She, Dongdong and Shah, Abhishek and Jana, Suman},
  booktitle={2022 IEEE Symposium on Security and Privacy (SP)},
  pages={2194--2211},
  year={2022},
  organization={IEEE}
}

@misc{0dayinthewild,
	author = {Google Project Zero},
	title = "0-days In-the-Wild",
	howpublished = "\url{https://googleprojectzero.github.io/0days-in-the-wild/}",
	year={2025}
}

@misc{weakptr,
	author = {cppreference},
	title = "std weak ptr",
	howpublished = "\url{https://en.cppreference.com/w/cpp/memory/weak_ptr.html}",
	year={2025}
}

@Book{Payer18SS3P,
  author  = {Mathias Payer},
  title   = {{Software Security: Principles, Policies, and Protection}},
  publisher = {HexHive Books},
  month   = {July},
  year    = {2021},
  edition = {0.37},
  url     = {http://nebelwelt.net/SS3P/},
}

@misc{bugcrowd_report,
	author = {bugcrowd},
	title = "Inside the Platform Vulnerability Trends Report",
	howpublished = "\url{https://mysecuritymarketplace.com/wp-content/uploads/2024/07/Inside-the-Platform-Vulnerability-Trends-Report.pdf}",
	year={2025}
}

@inproceedings{dullien2018security,
  title={Security, Moore’s law, and the anomaly of cheap complexity},
  author={Dullien, Thomas},
  booktitle={presentation at Conference on Cyber Conflict, May},
  year={2018}
}

@misc{chrome_full_rce,
	title = "ipcz bug can allow renderer duplicate browser process handle to escape sandbox",
  author={Micky},
  howpublished = "\url{crbug.com/412578726}",
  year={2025}
}

@misc{miracle_ptr_bypass,
	title = "MiraclePtr bypass due to PtrCount overflow",
  author={Micky},
  howpublished = "\url{crbug.com/340122160}",
  year={2025}
}

@misc{pwn2own25berling,
	title = "Pwn2Own Berlin 2025: Day Three Results",
  author={trendmicro},
  howpublished = "\url{https://www.zerodayinitiative.com/blog/2025/5/17/pwn2own-berlin-2025-day-three-results}",
  year={2025}
}

@inproceedings{kim2025logs,
  title={Logs In, Patches Out: Automated Vulnerability Repair via $\{$Tree-of-Thought$\}$$\{$LLM$\}$ Analysis},
  author={Kim, Youngjoon and Shin, Sunguk and Kim, Hyoungshick and Yoon, Jiwon},
  booktitle={34th USENIX Security Symposium (USENIX Security 25)},
  pages={4401--4419},
  year={2025}
}

@inproceedings{nong2025appatch,
  title={$\{$APPATCH$\}$: Automated adaptive prompting large language models for $\{$Real-World$\}$ software vulnerability patching},
  author={Nong, Yu and Yang, Haoran and Cheng, Long and Hu, Hongxin and Cai, Haipeng},
  booktitle={34th USENIX Security Symposium (USENIX Security 25)},
  pages={4481--4500},
  year={2025}
}

@inproceedings{zhang2022program,
  title={Program vulnerability repair via inductive inference},
  author={Zhang, Yuntong and Gao, Xiang and Duck, Gregory J and Roychoudhury, Abhik},
  booktitle={Proceedings of the 31st ACM SIGSOFT International Symposium on Software Testing and Analysis},
  pages={691--702},
  year={2022}
}

@inproceedings{shariffdeen2021concolic,
  title={Concolic program repair},
  author={Shariffdeen, Ridwan and Noller, Yannic and Grunske, Lars and Roychoudhury, Abhik},
  booktitle={Proceedings of the 42nd ACM SIGPLAN International Conference on Programming Language Design and Implementation},
  pages={390--405},
  year={2021}
}

@article{gao2021beyond,
  title={Beyond tests: Program vulnerability repair via crash constraint extraction},
  author={Gao, Xiang and Wang, Bo and Duck, Gregory J and Ji, Ruyi and Xiong, Yingfei and Roychoudhury, Abhik},
  journal={ACM Transactions on Software Engineering and Methodology (TOSEM)},
  volume={30},
  number={2},
  pages={1--27},
  year={2021},
  publisher={ACM New York, NY, USA}
}

@misc{zeroclick_pixel,
	author = "Natalie Silvanovich",
	title = "A 0-click exploit chain for the Pixel 9 Part 1: Decoding Dolby",
	howpublished = "\url{https://projectzero.google/2026/01/pixel-0-click-part-1.html}",
	year={2026}
}

@misc{zeroclick_imessage,
    author = {Bill Marczak and John Scott-Railton and Bahr Abdul Razzak and Noura Aljizawi and Siena Anstis and Kristin Berdan and Ron Deibert},
    title  = {NSO Group iMessage Zero-Click Exploit Captured in the Wild},
    howpublished = {\url{https://citizenlab.ca/research/forcedentry-nso-group-imessage-zero-click-exploit-captured-in-the-wild/}},
    year   = {2023}
}

@misc{oneclick_android,
	author = "New York Times",
	title = "Read the Intellexa Pitch on Its Spyware Tool",
	howpublished = "\url{https://www.nytimes.com/interactive/2022/12/08/us/politics/intellexa-commercial-proposal.html}",
	year={2022}
}
\bibliographystyle{tmlr}

\appendix

\end{document}